## Authors

Jonas Birgersson[1], Dr. Marc A. Weiss[2], Dr. Jimmy Chen[3], Dr. Daniel Kammen[4], Dr. Tomas Kåberger[5], Dr. Franklin Carrero-Martínez[6], Dr. Joakim Wernberg[7], Dr. Michael Menser[8], Dr. Newsha K. Ajami[9].

[1] EnergyNet Task Force; ViaEuropa Sverige AB, Lund, Sweden.

[2] Global Urban Development; University of California, Berkeley, United States.

[3] Stanford University, Precourt Institute for Energy, Doerr School of Sustainability, Stanford, United States.

[4] Johns Hopkins University, Bloomberg Distinguished Professor of Energy and Climate Justice, Department of Civil and Systems Engineering, and Paul Nitze School of Advanced International Studies, Baltimore and Washington DC, United States.

[5] Chalmers University of Technology, Department of Technology Management and Economics, Gothenburg, Sweden.

[6] National Academies of Sciences, Engineering, and Medicine, Global Sustainability and Development, Washington DC, United States.

[7] The Swedish Entrepreneurship Forum; Lund University, Department of Technology and Society, Sweden.

[8] Brooklyn College and City University of New York (CUNY) Graduate Center; NYC Climate Justice Hub and Public Power Observatory, New York, United States.

[9] Lawrence Berkeley National Laboratory, Earth and Environmental Sciences Area, Berkeley, United States.

Author Contributions

J.B. conceived the main ideas and led the writing. M.W. coordinated research and manuscript preparation. J.C. contributed with important ideas and writing. D.K. provided critical feedback and edits. T.K. provided critical feedback and edits. F.C-M. provided critical feedback and edits. J.W. provided critical feedback and edits. M.M. provided critical feedback and edits. N.A. provided critical feedback and edits.

Corresponding Author

Jonas Birgersson

Email: jonas.birgersson@viaeuropa.net

Address: ViaEuropa Sverige AB, Winstrupsgatan 1, 222 22 Lund, Sweden

# EnergyNet Explained: Internetification of Energy Distribution

*A How-To Guide for a 21st-century Energy System*

*Abstract*

*In developing EnergyNet we have leveraged and are extending lessons from telecom's shift from a centralized, circuit-switched phone system to decentralized, packet-switched data networks. EnergyNet utilizes: 1) an Energy Router that enforces galvanic separation and utilizes software-controlled energy flows over a DC backplane, 2) Energy Local/Wide Area Networks (ELAN/EWAN) based on DC microgrids that interconnect through an open Energy Protocol (EP), and 3) a control plane comprised of the Energy Router Operating System (EROS) and EP Server which is managed at operator scale through an Energy Network Management System (ENMS). We distinguish the architectural contribution (Tier-1: components, interfaces, operating model) from expected outcomes contingent on adoption (Tier-2). The latter includes local-first autonomy with global interoperability, near-real-time operation with local buffering, removal of EV-charging bottlenecks, freed grid capacity for data centers and industrial electrification, as well as a trend toward low, predictable, fixed-cost clean energy. Evidence from early municipal demonstrators illustrates feasibility and migration paths. The contribution is a coherent, open, and testable blueprint for software-defined, decentralized energy distribution, aligning power-systems engineering with networking principles and offering a practical route from legacy, synchronous grids to resilient, digitally routed energy distribution systems.*

# 1. Executive Summary: A New Grid Architecture for the 21-century

This paper introduces EnergyNet, an Internet-inspired architecture for electricity distribution, specifying its components, interfaces, and operating model.

EnergyNet is a modular, open architecture for energy distribution, with these key elements:

- the Energy Router (galvanic separation, DC backplane, variable-voltage ports).
- the control plane (the new open source Energy Protocol and Energy Router OS + EP-Server).
- ELAN/EWAN boundaries and the Energy Protocol for interdomain negotiation.
- the operator model (ENMS + BSS/OSS/eTOM alignment).
- deployment patterns (e.g., new DC microgrid infrastructure and co–deployment with fiber/LTDH), security, and governance assumptions.

The existing grid or Plain Old Grid System (POGS), while a technological marvel of its time, was built for the needs and technologies of the 20th century. Its centralized, rigid, and increasingly complex structure has now become a bottleneck, slowing progress at a moment when acceleration is urgently needed. Specifically, transmission to distribution linkages and the operational dynamics of distribution grids are not yet on a trajectory to become smart, adaptive systems capable of serving human, commercial, and industrial demands.

EnergyNet is a transformative system architecture, not a single technology, product, or vendor. By combining modular power electronics, software-defined networks, and new open protocols, it turns the distribution layer of the grid into a flexible, decentralized network of networks that can adapt to local needs and national demands. Here are some key differences between POGS and EnergyNet:

- • From centralized to distributed energy systems, built for scalability and replicability.
- • From analog control to digital coordination with software-defined routing.
- • From closed proprietary systems to open source, enabling democratization of energy.

Drawing on Europe's deregulation successes and energy community reforms, we argue that, like mobile and broadband before, the next infrastructure wave primarily can be funded by market actors, and does not rely on government subsidies or monopoly charges.

This paper introduces the EnergyNet model, an Internet-style architecture for energy distribution, and provides a practical how-to for homes, buildings, communities, and urban/regional deployment. We define components and interfaces (Energy Router, ELAN/EWAN, Energy Protocol, EROS/EP-Server, ENMS), reference early demonstrators, and outline a scale-up roadmap. The aim is to enable replication and extension, inviting further research, pilots, and open implementation.

Our paper is written for those who want to move from concept to implementation, and lead the way in building a new grid for the 21st century.

Acknowledgments

The authors acknowledge valuable input from colleagues, industry experts, and support from OpenAI's ChatGPT in manuscript drafting, editing, and refinement. The authors bear full responsibility for all content presented.

## 2. Introduction & Purpose

Across the globe, the energy transition is facing a paradox. On one hand, renewable energy technologies like solar panels, batteries, and electric vehicles are becoming more affordable and available than ever before. On the other hand, our legacy energy distribution infrastructure – the traditional Alternating Current (AC) grid, or Plain Old Grid System (POGS) – is increasingly becoming a bottleneck. The result is that in many areas, renewable projects are being delayed or even canceled due to a lack of available grid capacity.

This is not a technology problem; it is an architecture problem. POGS was designed for one-way flows of power from centralized plants to passive consumers. It was never built to support dynamic, bidirectional flows between thousands of producers and consumers at the edge of the network.

The EnergyNet model, grounded in open standards and software-defined networks, offers a blueprint that any real estate owner or developer, city, region, or country can adopt and adapt, regardless of their starting point. It allows for rapid, market-driven deployment at scale, similar to how mobile networks and broadband scaled globally. Inspired by how the Internet transformed telecommunications, the EnergyNet model enables energy flows to become both adaptable and scalable, using software-based coordination across a "network of networks", corresponding to the decentralized infrastructure of the Internet.

### Framing the Problem: A Net-Head Perspective on the Traditional AC Grid

This work began with growing frustration at today's traditional power grid. Once a marvel of human ingenuity that powered the early phase of electrification, it transformed societies across the 20th century. But what was revolutionary is now showing its age. The centralized, rigid, and complex distribution architecture has become a limiting factor. As new technologies proliferate and demand patterns shift, complexity and costs escalate while relief remains out of reach. The legacy grid has taken us far; its foundational structure now stands in the way of the flexibility, resilience, and innovation the future requires.

A similar moment arrived for fixed telephony (POTS -- Plain Old Telephone System) in the late 1990s, when traffic shifted from short human calls to long modem sessions. The system strained and a fundamental architectural debate ensued. Two camps emerged: "Bell-heads," who pushed for bigger, smarter central switches, and "Net-heads," who argued for a simpler, decentralized architecture that pushed control to the edge. David Isenberg crystallized the Net-head viewpoint that the new "stupid network is a very smart idea" in an impactful article in 1997 characterizing it as a deliberately underspecified, abundant, packet-based fabric ("bits in, bits out") that unleashes innovation at the edge [1].

In 1998, in Lund, Sweden, a small team led by fellow Net-head Jonas "Birger" Birgersson built a proof-of-concept along these lines. By 1999 [2], [3], [4], [5], [6], Bredbandsbolaget deployed Ethernet-to-the-Home at national scale using commodity data-network gear to deliver 10/10 Mbps Internet for a low fixed price of 200 SEK (~USD 20) per month. When the incumbent Telia matched the offer later that year, the architectural debate was effectively settled; the principles behind the digitally innovative "stupid network" eventually became the global broadband standard.

The authors believe that we now face an analogous inflection point in energy. Just as the shift from circuit-switched POTS to packet-switched Internet required rethinking the fabric of telecommunications, POGS requires structural change: a distribution architecture that is modular, software-defined, and adaptable by design, with local autonomy and policy-based interconnection between domains.

The next sections introduce the EnergyNet model and its coordination logic, showing how this system-level shift can be implemented now with existing technologies, provided we apply a coherent EnergyNet architecture approach.

## 3. Why EnergyNet?

For a century, electricity was generated at a small number of large power plants and pushed outward over one-way, centrally coordinated networks. That architecture -- radial feeders, synchronous coupling, and centralized protection -- was optimized for predictable, top-down power flows.

The transition underway is the opposite. Generation is decentralizing; rooftop PV, batteries (including EVs), heat pumps, and flexible loads live at the edge. Power flows become bidirectional and intermittent, and coordination must happen locally and quickly. The legacy grid was not built for this: capacity fills, protection coordination breaks down, frequency/voltage control gets challenged, and single points of failure at substations and feeders reveal structural fragility.

### 3.1 If We Were Building the Grid Today

Starting from a clean slate, a modern system would be local-first and digitally coordinated: microgrids with power-electronics frontiers (galvanic separation), software-defined energy flows, open protocols, local buffering/storage (near-real-time instead of strict real-time), and policy-based interconnection between domains. In short, a network of networks, simple and robust at its core, through dynamic capacity, resilience, and scale, by adding ports and nodes as needed.

### 3.2 EnergyNet: From Scarcity to Abundance

EnergyNet is a blueprint. It replaces a centralized, hierarchically managed system with a modular, software-defined, interoperable architecture that lets communities produce, store, share, and trade energy locally, while interconnecting safely with the wider grid. By pushing intelligence to the edge and making interconnection simple, negotiated, and dynamic, EnergyNet turns the present bottleneck into a platform for growth: from scarcity to abundance, with lower and more predictable costs, greater resilience, and a faster path to electrification at scale.

In this paper, by applying principles of decentralization, openness, and local empowerment, we explore why the new grid that EnergyNet enables is so critically important for society. We will demonstrate how this new architecture could lead to an abundance of affordable clean energy, more resilient distribution, and improved energy independence.

EnergyNet reimagines energy infrastructure by applying the principles of the Internet to energy distribution:

- Local Generation: Solar panels and small-scale renewables generate energy where it is consumed, reducing the need for long-distance transmission.
- Local Storage: Home batteries and EVs store excess generation for local use when local production is not available or demand spikes.
- Local Sharing: DC microgrids and simple energy routing dynamically balance loads, generation, and storage, reducing peak demand stress on the broader grid.

This local, decentralized approach lowers operational costs (OPEX) and capital expenditure (CAPEX) per delivered kWh, enabling communities to move towards predictable, low fixed costs for the green energy they generate, store, and share locally.

| Dimension | POGS | EnergyNet |
|---|---|---|
| Architecture | Centralized, radial AC; one big synchronous system. | **Decentralized** ELAN/EWAN; Energy Routers with DC backplane. |
| Interconnection | Synchronous, hard-wired coupling. | **Digitally negotiated** with **galvanic separation** on each port. |
| Control Plane | SCADA + manual ops, vendor silos. | **Software-defined**, Open Standard; **Energy Protocol (EP).** |
| Flow Model | Continuous real-time only. | **Near-real-time with local buffering**; "some power" flexibility. |
| Economics | Central CAPEX, long payback; require monopoly for investment. Risk for stranded assets. | **Pay-as-you-grow**, market-driven CAPEX tied to local value; interoperable with and complementary to the existing grid. |
| Scaling | Slow, sequential reinforcements; interconnection queues; single-path constraints; timescale in years. | **Fast, parallel scaling**; modular, demand-led build-out; multi-path routing; can reuse existing infrastructure or build new, timescale in weeks. |

Fig. 1. POGS versus EnergyNet, architectural, operational, and economic contrasts.

## 3.3   The Hidden Power of Homes

The housing sector is one of the largest, most overlooked levers in the energy transition. In the United States [7], residential buildings account for approximately 38% of total electricity consumption annually, representing over 1.5 trillion kilowatt-hours in 2022 alone. In the European Union, same year [8], households consumed an estimated 29% of final electricity consumption, with variations depending on climate, heating sources, and building efficiency.

While homes are often seen as passive consumers, this immense base of distributed demand can be transformed by EnergyNet into an active, flexible grid resource. Rooftop solar, home batteries, and EVs with bidirectional charging turn all homes into dynamic nodes for renewable energy generation, storage, and sharing in a fully decentralized system that still is interoperable with the traditional grid. By connecting them through open standards like the Energy Protocol, the EnergyNet architecture unlocks this latent capacity, not just for individuals to benefit, but as a cornerstone of national energy independence and resilience.

Every building, district, and municipality has a huge untapped potential: rooftops, space for batteries, and the ability to build new smart microgrids. When orchestrated through open standards and software-defined networks with dynamic sharing, these form the foundation of a new energy sovereignty. From the local perspective it is more important to determine how much energy will be needed by the community, rather than just calculating the amount of kWh that can be generated and stored.

## 4. EnergyNet: the Internetification of Energy Distribution

Just like the telecom industry once relied on the POTS, the world's energy infrastructure still remains dependent on the POGS. Like POTS, the traditional energy grid was groundbreaking at its inception but has become a rigid, outdated infrastructure incapable of adapting to today's technological innovations and decentralized renewable energy solutions.

EnergyNet is a new grid architecture. Similar to how Fiber-to-the-Home (FTTH) and Fiber-to-the-Building (FTTB) broke the bottleneck of the POTS, EnergyNet can now break the bottleneck of the POGS. We describe the path forward in practical terms on how to build the new grid for the 21st-century.

### How the War in Ukraine showed the Vulnerability of the Old Grid Architecture

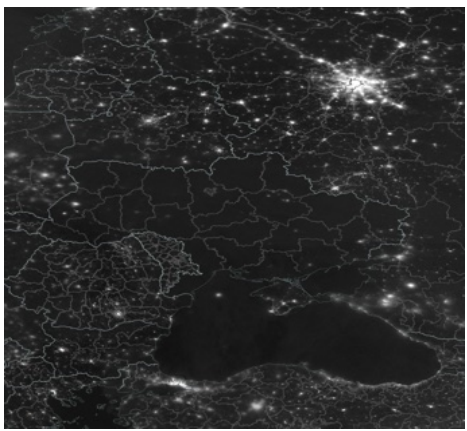

Fig. 2. NASA satellite image of from November 24, 2022. Source: NASA/NOAA Black Marble [9], illustrating the national blackout effect in Ukraine.

The November 2022 missile attacks on Ukraine's energy infrastructure vividly illustrated the critical vulnerability of centralized grid systems, referred to as POGS. Despite Ukraine's vast geography and extensive energy infrastructure, a relatively small number of targeted missile strikes on crucial facilities resulted in severe and widespread blackouts.

A handful of missile hits had a disproportionately large impact, exacerbated significantly by necessary protective measures implemented by Ukraine's Transmission System Operator, Ukrenergo [10]. Despite heroic restoration efforts by Ukrainian engineers and energy workers, protective measures included preemptive blackouts to safeguard the grid from catastrophic cascading failures, further magnifying the societal and economic consequences.

The stark NASA satellite image (Fig. 2) from November 24, 2022, powerfully captured this impact, showing Ukraine plunged into near-total darkness in contrast to neighboring countries still illuminated. This event highlights the fundamental fragility inherent in centralized energy systems and underscores the urgent need for decentralized resilient energy solutions capable of maintaining local power supplies despite targeted disruptions.

### 4.1  Demonstrating Internet Resilience through Decentralized Architecture

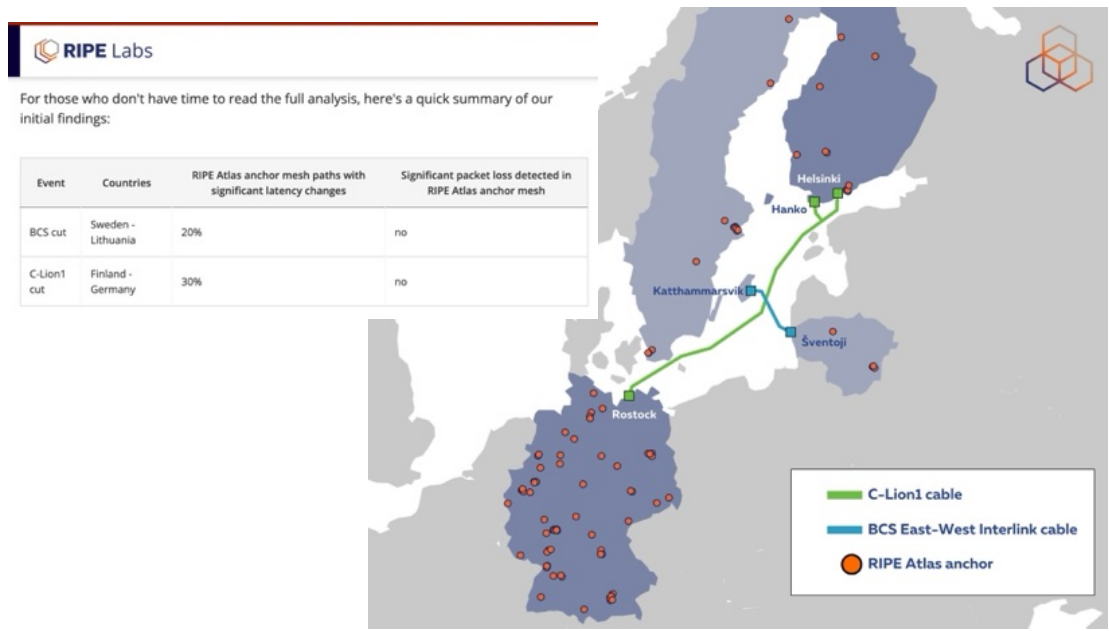

Fig. 3. Source: RIPE Labs [11]; Map demonstrates the Internet's resilience, with its "no- single-point-of-failure" architecture for critical infrastructure that already is a practical fact.

The Baltic Sea cable disruptions in November 2024, involving the simultaneous cuts of the C-Lion1 (Finland–Germany) and BCS East-West Interlink (Sweden–Lithuania) submarine cables, clearly illustrated the resilience of the Internet's decentralized architecture. Despite the severity and strategic timing of these disruptions, RIPE Labs' comprehensive analysis using the RIPE Atlas network revealed minimal negative impact.

Specifically, RIPE Labs [11] found no significant packet loss across their monitoring network, indicating that data flows swiftly rerouted around the damaged cables. Although 20% to 30% of Internet paths experienced minor latency increases, the majority remained unaffected, underscoring the Internet's built-in redundancy and adaptive routing capabilities.

This incident sharply contrasts the vulnerabilities exposed in traditional centralized infrastructures, such as electricity grids, during similar targeted attacks. It highlights the superior resilience of distributed, redundant, and dynamically adaptive network architectures. The Internet's robust response to the Baltic Sea cable cuts thus provides a proven architectural template, reinforcing the rationale for adopting similarly resilient decentralized approaches, such as EnergyNet, in critical infrastructure systems.

## 4.2 System Architecture Overview

A Software-Defined Energy Distribution Layer

The core innovation behind EnergyNet is architectural: it decouples local energy systems from the constraints of the legacy grid and enables flexible, digitally orchestrated distribution using modern digital technology. Like the Internet did for information, EnergyNet introduces a software-defined routing layer for electricity.

## 4.3 Internet as an Architectural Model for EnergyNet

To understand EnergyNet's approach to energy distribution, consider how the Internet itself is structured. The Internet is not one single physical network; it's a hierarchy of logically defined networks, each optimized for specific tasks: LAN, WAN, and global Internet.

### 4.3.1 Local Area Network (LAN)

A LAN is the local network, such as your home or office network, which connects your devices. It's defined logically, not geographically, by local control and high-performance internal connectivity. In energy terms, this corresponds to a single building or a neighborhood microgrid, which manages local energy flows autonomously, prioritizing resources according to where and when energy is needed.

### 4.3.2 Wide Area Network (WAN)

A WAN connects multiple LANs, linking networks logically rather than by simple geography. An example is an Internet Service Provider's network, which connects thousands of individual home and business LANs across a city, country, or globally. In EnergyNet, the equivalent is a district or citywide energy distribution network, coordinating energy flow across multiple neighborhoods or districts, dynamically balancing energy resources. Please note that WANs are logical, not primarily geographic concepts.

### 4.3.3 The Internet

Above WANs sits the global Internet — an interconnected system of independent WANs, each managed as Autonomous Systems (AS). This global layer is where WANs communicate using standardized protocols like Border Gateway Protocol (BGP), allowing decentralized yet coordinated global routing of data. In the EnergyNet analogy, this represents the overall interconnected energy network — coordinated globally yet managed independently and locally. Just as BGP provides a secure, open, and robust mechanism for global data routing, the EnergyNet's open protocols allow similar dynamic coordination, based on different policy inputs such as pricing or priority for resilience, and enable large-scale sharing of energy.

This layer model, devices within LANs, LANs interconnected by WANs, and WANs coordinated globally by Internet standards, maps directly onto EnergyNet's architecture. Each local energy router (LAN equivalent) autonomously manages its own resources, clusters of energy routers (WAN equivalent) dynamically negotiate energy exchanges, and a global coordination protocol (BGP equivalent) ensures flexible, secure, and scalable orchestration within the EnergyNet, and digital interaction with the traditional grid operators.

Thus, the Internet's proven logical structure provides a roadmap for how EnergyNet can redefine energy distribution: decentralized, software-defined, and coordinated dynamically, at every level.

## 4.4 Legacy Grid (POGS) vs EnergyNet: A Structural Shift

The traditional AC grid is based on a centralized, top-down architecture designed for unidirectional flow: power plants generate, transmission lines carry, and end-users consume. This design made sense when electricity generation was scarce, centralized, and predictable.

In contrast, today's energy landscape is becoming more decentralized, dynamic, and digital. Homes, buildings, and vehicles can now generate, store, and share energy. But the legacy grid – POGS -- was never built to accommodate this. EnergyNet addresses this mismatch by offering a parallel routing layer that operates alongside the existing grid, providing:

- Local autonomy with global interoperability.
- Bidirectional, prioritized energy flows.
- Digital control via open coordination protocols.
- Firewalled interactions with the legacy grid.

This is not a replacement of the legacy grid; it's a layered new integrated architecture. EnergyNet adds intelligence, modularity, and programmability to the edge of the grid, allowing energy to be routed, priced, and prioritized in near-real-time.

## 4.5 Core Concepts: EnergyNet Digital Architecture

EnergyNet introduces a new, digital architecture for energy distribution. It decouples local energy systems from the constraints of traditional centralized grids; instead it enables dynamic, flexible energy flows. Inspired by the Internet's architecture, EnergyNet applies open standards, software-defined control, and decentralized coordination to create a robust, responsive energy infrastructure. EnergyNet's architecture revolves around three fundamental principles:

- Local autonomy with global interoperability.
- Bidirectional and dynamic energy routing.

- Software-defined coordination and open protocols.

## 4.6 From Internet to EnergyNet

EnergyNet's layered architecture mirrors the layered model of the Internet, clarifying roles, boundaries, and interactions at different network levels:

- LAN → ELAN (Energy Local Area Network): Manages energy flows within buildings or neighborhoods, enabling immediate local optimization and autonomy.
- WAN → EWAN (Energy Wide Area Network): Connects multiple ELANs, facilitating dynamic energy exchanges, load balancing, and regional coordination.
- Internet → EnergyNet (Global Network): Links EWANs together on a national or international scale, coordinating energy distribution through standardized protocols and decentralized management.

Boundaries between ELAN, EWAN, and EnergyNet global networks are clearly defined and operate similarly to Internet routing hierarchies, utilizing internal coordination protocols analogous to OSPF and external protocols analogous to BGP.

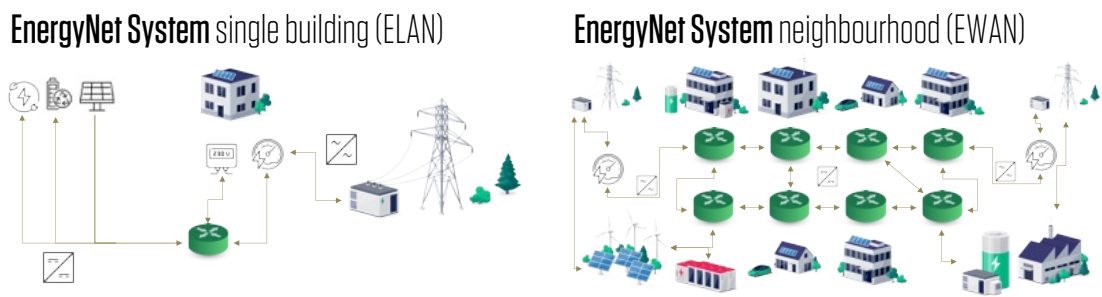

Fig. 4a (left). EnergyNet with ELAN in a single building illustrating the core position of the Energy Router. Fig. 4b (right). EnergyNet with EWAN for a neighborhood with ring architecture with an additional interconnection.

## 4.7 Core Principles: The Firewall

EnergyNet solves frequency stability challenges in the traditional grid with digitally controlled galvanic "Firewall" functionality built into every Energy Router.

In the Alternating Current grid, frequency stability is critical. It is sometimes disturbed by unpredicted losses of power lines or large thermal power plants. When there were many synchronous generators in the system these contributed with short term stabilizing power. In modern systems you need electronic components to handle failures of remaining large thermal units or power lines. The batteries and electronics of the EnergyNet can contribute with stabilizing services.

### 4.7.1 Software Controlled Galvanic Separation as a Solution

EnergyNet addresses this challenge by introducing a digitally controlled "galvanic separation" functionality, a deliberate electrical isolation between local energy resources and the traditional grid, implemented via power electronics in the Energy Router.

Unlike a direct electrical connection, galvanic separation creates a clear digital boundary. Power flows across this boundary when, and only when, both sides explicitly agree, using a predefined digital negotiation process managed by software. This makes it possible to support the grid, while it will never be disturbed.

This approach is comparable to how the clearly defined border between LAN and WAN works on the Internet. In a simple example, the device managing the border is often called a Gateway. This unit typically contains at least these three functions:

- Switching (Layer 2): Allowing devices within the local network (LAN) to communicate efficiently at high speed.
- Routing (Layer 3): Managing traffic between the local network (LAN) and external networks (WAN/Internet) using IP addressing.
- Firewall & NAT Functions: Implementing basic security measures — Network Address Translation (NAT) and firewall rules — to protect the local network from unauthorized external access.

The Gateway device has two distinct sides:

LAN Side (Local):

- Uses private IP addresses (e.g., 192.168.x.x or 10.x.x.x).
- Manages communication internally within your home or office.
- Ensures all devices can communicate seamlessly within the local network.

WAN Side (Global):

- Assigned a global (public) IP address by your ISP.
- Manages external communication, connecting your LAN to the wider Internet.

Between these two sides, the Gateway typically implements Network Address Translation (NAT):

- NAT Functionality: Converts private local addresses into a single, publicly routable IP address, allowing multiple devices to share a single external connection.
- Traffic Flow Control: Communication between LAN and WAN sides occurs only when explicitly agreed upon — initiated by devices on the LAN side or configured through rules. No data crosses unless both sides confirm compatibility and security.

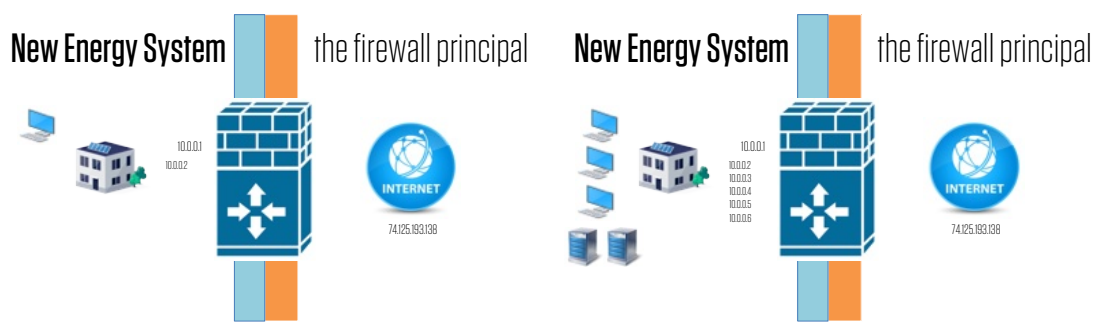

Fig. 5a (left). Illustration of the Firewall principle: local side is independent from global side but still interconnected. Fig. 5b (right). New devices are added to local side without need for more resources from the global side.

A gateway provides basic firewall principles, in the sense that it acts as a clear, managed boundary controlling traffic flow between two distinct networks. Specifically, it:

- Blocks unsolicited external traffic: By default, the gateway allows local traffic to initiate connections outward but prevents unsolicited inbound connections from the WAN side.
- Manages rules and permissions: Like a simple firewall, it allows users to define rules about what type of communication is permitted, blocked, or selectively forwarded.
- Ensures controlled interoperability: Traffic only crosses from WAN to LAN if explicitly permitted, protecting the local network from unwanted or unauthorized external communication.

The Gateway demonstrates the Firewall principles as a protective boundary between local and external networks, ensuring both independence and interoperability.

Each side (LAN and WAN) operates independently:

- If disconnected from the global (WAN) side, the LAN side continues functioning autonomously, allowing local devices to communicate without interruption.
- The Gateway's internal router functionality ensures local traffic is seamlessly managed even without WAN connectivity.

In short, a Gateway with integrated routing serves as a controlled and intelligent boundary, enabling secure, deliberate interaction between local and global networks, while preserving full local functionality in isolation.

### 4.7.2 Benefits of Digitally Managed Galvanic Separation

This digital approach provides multiple advantages:

- Frequency Stability: Local renewable energy resources can operate independently, without directly impacting the main grid's frequency stability.
- Controlled Interaction: Energy exchanges occur only when beneficial to both sides, preventing local fluctuations from cascading into the wider network.
- Dynamic Flexibility: Digital rules can be easily adapted to changing conditions or policies, allowing dynamic balancing and negotiation of energy transfers.

In short, galvanic separation through digitally managed power electronics not only solves the frequency stability problem introduced by intermittent renewables but also introduces a safer, more flexible, and robust method of coordinating energy distribution across different layers of the grid.

## 4.8 Core Principles: Near-Real-Time Power Distribution

EnergyNet's robust and scalable architecture enables it to function in "near-real-time" instead of real-time. A core principle underlying EnergyNet's advanced architecture is the transition away from a strict real-time model, like the traditional power grid or the Plain Old Telephone System (POTS), towards a more robust and scalable near-real-time system inspired by the Internet's packet-switched network model.

Historically, the electrical grid required continuous, instantaneous balancing of supply and demand, with any mismatches immediately harming grid stability. Similarly, the legacy telephone system required dedicated circuits and real-time connections, leaving no room for delays or disruptions.

EnergyNet adopts a fundamentally different approach, using local energy storage and digitally managed buffering as analogs to the Internet's data caches and packet buffers. Instead of instantaneous power flows, energy is locally buffered, stored, and digitally routed based on software-defined priorities and negotiated exchanges. This "packetized energy" approach provides:

- Enhanced Robustness: Short-term storage and buffering make the system inherently tolerant to brief interruptions or imbalances.
- Simplified Management: Near-real-time management dramatically reduces complexity by eliminating the need for instantaneous global balancing.
- Scalability: Like packet switching in telecommunications, this architecture scales easily. New resources and storage can be seamlessly added, increasing capacity without complicating management.

By moving from strict real-time to near-real-time operation, EnergyNet achieves the same benefits that packet switching brought to telecommunications: a dramatically simpler, more resilient, and easily scalable system, perfectly suited for the diverse, distributed energy resources of the 21st-century.

## 4.9 Core Principles: The Value of Some Power vs Blackout

In the traditional electrical grid, power delivery is fundamentally binary, either fully operational at 100% or experiencing a complete blackout at 0%. A significant drop in frequency or voltage typically triggers cascading failures, causing the grid to collapse entirely. This makes the traditional grid vulnerable and brittle, leaving critical infrastructure completely powerless during major disruptions.

EnergyNet fundamentally changes this dynamic by adopting a digitally coordinated, modular, and locally autonomous architecture. Instead of an "all-or-nothing" scenario, EnergyNet can maintain partial, but highly targeted, energy delivery under nearly any circumstance.

Key aspects of this resilience principle include:

- Priority-Based Energy Delivery:

EnergyNet continuously identifies and prioritizes essential equipment and critical services (such as routers, communications equipment, medical devices, or emergency lighting). In a disruption, the system automatically reallocates limited available power, ensuring the highest priority assets remain operational.

- Extended Partial Operation:
  By intelligently managing local energy storage and generation resources, EnergyNet can sustain partial operations for extended periods, even if grid connections or higher-level communication networks (such as global cloud services) are disrupted. The local digital control within Energy Routers maintains robust command and coordination capabilities independently.

- Dynamic Adaptation and Local Autonomy:
  EnergyNet's decentralized design allows each local node to independently reconfigure energy flows based on real-time conditions. Even if completely isolated from external networks or the traditional grid, EnergyNet dynamically reallocates and reshapes available resources to optimize survival and operational effectiveness.

This principle, ensuring "some power" rather than "no power", represents a transformative improvement in energy resilience. EnergyNet ensures critical digital infrastructure, communication capabilities, and essential local resources can remain functional far longer and more reliably than traditional grid infrastructure would allow, especially during crises or extended disruptions.

## 4.10 Key EnergyNet Software Component: The Energy Protocol

The Energy Protocol is an open standard communication framework at the heart of EnergyNet. It enables Energy Routers and networks to securely negotiate, coordinate, and exchange energy data, ensuring interoperability across different systems and vendors. Similar to Internet protocols like TCP/IP, the Energy Protocol enables reliable, transparent, and secure energy routing decisions [12].

## 4.11 Energy Router

At the core of EnergyNet's digital distribution architecture is the Energy Router, a power-electronics device designed to manage energy flows dynamically, software-defined, and safely. Like data routers in communication networks, the Energy Router directs electric power precisely where and when it's needed, enabling a flexible, responsive, and resilient energy system.

### 4.11.1 Variable Voltage Capability of Energy Router Ports

The Energy Router's ports can dynamically manage and vary the voltage level of the DC power being sent or received. Depending on system needs, local conditions, or connected equipment, these ports operate flexibly within defined voltage ranges, such as:

- 150 to 800 V DC (typical for many local storage solutions and EV charging).
- 150 to 1500 V DC (extended range suitable for larger systems, utility-scale storage, or advanced infrastructure requirements).

This variable voltage capability allows the router to precisely control power delivery, optimize efficiency, and ensure compatibility with a wide range of local generation, storage, and consumption devices.

### 4.11.2 Rapid Improvements in Price-Performance for Bidirectional Converters

In recent years, bidirectional power electronics -- especially AC/DC and DC/DC converters used in electric vehicles [13], battery storage systems, and local renewable energy integration -- have experienced dramatic advances in both performance and affordability. Driven by the adoption of advanced semiconductor technologies [14], particularly silicon carbide (SiC) and gallium nitride (GaN), converters now achieve significantly higher efficiency (often exceeding 98%), greater power density, and more compact sizes.

These advancements have already sharply reduced costs per kilowatt (kW), with the price-performance ratio improving at rates comparable to the early stages of Moore's Law in computing. Industry experts anticipate that this trend will accelerate further, with continuous improvements expected over the next decade. Increasing production volumes, particularly driven by rapid electric vehicle adoption and widespread deployment of storage systems, will further drive down costs through economies of scale.

Looking ahead, future bidirectional converters are projected to become even smaller, more efficient, and cost-effective. Enhanced integration, simplified designs, and standardized modular architectures will lead to widespread adoption of high-performance, low-cost converters across all areas of energy management. These technological and economic trends will be a great driver for further accelerating the global transition towards dynamic, digitally coordinated energy networks, such as EnergyNet.

### 4.11.3 Dual Energy Supervisors: Built-In Redundancy for Critical Coordination

To ensure high availability and avoid any single point of failure, each Energy Router is typically equipped with two Energy Supervisors, redundant controller units that work together to provide uninterrupted coordination and operation.

Each Energy Supervisor runs the core digital systems that define the router's behavior:

- EROS (Energy Router Operating System): Responsible for managing internal port-to-port energy routing logic, local prioritization, and switching decisions across the DC backplane.
- EP-Server (Energy Protocol Server): Manages secure negotiation and coordination with other Energy Routers and with the EnergyNet Operator using the open Energy Protocol and advanced Energy Network Management System (ENMS).

These dual supervisors operate in an active-passive or active-active failover mode, ensuring that if one module fails or requires maintenance, the other immediately takes over without disrupting energy flows or control logic. All configurations, energy routing tables, and protocol states are mirrored in real time.

This architectural redundancy mirrors the proven design practices found in telecom-grade and cloud infrastructure systems and is key to the EnergyNet mission of building self-healing, highly resilient, and digitally autonomous microgrids.

By combining this dual-controller design with modular power ports and galvanic isolation, the Energy Router provides a dependable foundation for next-generation energy distribution, especially in mission-critical, community-scale deployments.

### 4.11.4 Internal DC Backplane: Enabling Intelligent Energy Routing

At the heart of the Energy Router is an internal DC backplane, functioning like a high-capacity internal "energy bus." This DC backplane connects all ports and provides a shared interface through which the Energy Router can, through software-defined commands, route energy from any input port (or ports) to any output port (or ports) depending on real-time demands and priorities.

### 4.11.5 Modular, Rack-Based Design of Energy Routers

Energy Routers follow a modular design philosophy inspired by the telecommunications industry, leveraging standard 19-inch telecom racks to simplify installation, scalability, and maintenance. Each Energy Router consists of individual converter modules, AC/DC or DC/DC, that can be added or replaced independently, allowing capacity to grow incrementally and cost-effectively as needed.

Thanks to significant improvements in power electronics (particularly compact, highly efficient GaN and SiC converters), modern converter modules are dramatically smaller and more powerful than previous generations. A single converter module often occupies as little as 1U (1.75 inches) of vertical rack space, sometimes supporting multiple ports in one module.

### 4.11.6 Capacity of a Standard 42U Rack

Given current industry developments, a standard 42U telecom rack can comfortably host a substantial number of Energy Router ports.

Typical module density:

- 1 or 2 ports per 1U module (common configuration); approximately 42 to 84 ports per full 42U rack.
- Advanced high-density modules (emerging): up to 4 ports per 1U module, potentially enabling more than 150 ports per rack.

In practical deployments, a reasonable, conservative estimate would be between 50–80 bidirectional ports per full 42U rack, ensuring ease of cable management, cooling, and maintenance.

This modular and scalable approach ensures Energy Routers can seamlessly adapt and expand to match evolving local energy requirements, maximizing flexibility, minimizing upfront investment, and future-proofing the energy infrastructure.

### 4.11.7 Modular Design Enables Highly Resilient Energy Infrastructure

The modular, rack-based architecture of Energy Routers, combined with significantly lower costs per port, enables an entirely new level of system resilience. Because individual converter modules and their associated ports are inexpensive and easy to add, it's now practical to design networks without single points of failure.

With this approach:

- Distributed Redundancy: Multiple redundant ports and converters can be installed at low cost. If one module or port fails, energy flows automatically reroute through alternative functioning ports.
- Simplified Maintenance: Modular components can be individually replaced without shutting down the entire system, significantly reducing downtime.
- Enhanced Reliability: By eliminating reliance on a single critical component, the overall system becomes highly resilient, ensuring continuous operation and robust local energy autonomy, even in case of individual failures.

In short, modularity and cost efficiency don't just improve scalability and economics, they fundamentally enhance reliability, resilience, and system-wide robustness.

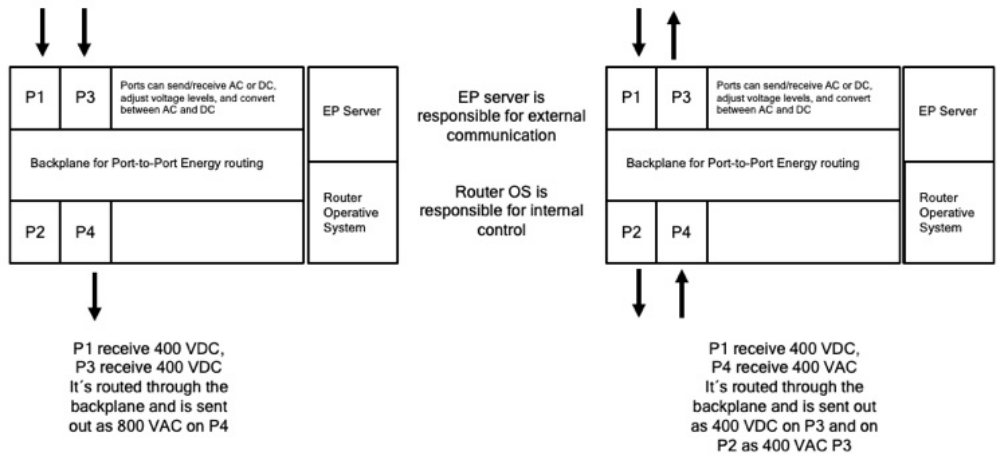

Fig. 6. Illustration of the main components of the Energy Router: ports, Operating System, and the EP-server. It also shows how electricity can be moved from any port to any combination of ports depending on the needs communicated over the Energy Protocol.

## Energy Router: four logic sides

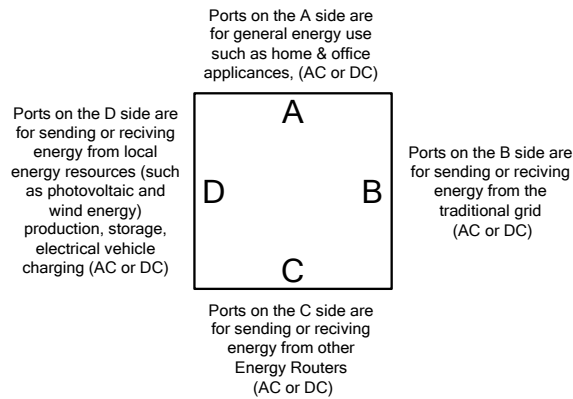

Fig. 7. Illustration of the four logic sides, same port but with different software logic applied based on what type of device is connected.

Acting as a digital gateway between ELAN/EWAN and the traditional grid, each router includes up to four different logic type of ports:

- Port A – AC port for local consumption (power to building or to tenants).
- Port B – AC port for traditional grid interconnection.
- Port C – DC port connecting to other Energy Routers for energy sharing.
- Port D – DC port connecting local energy resources (solar PV, batteries, EV chargers, etc.).

### 4.12 EnergyNet Operator

EnergyNet Operators perform roles analogous to Internet Service Providers (ISPs). They coordinate energy distribution, manage regional networks, and ensure reliable system-wide operations. Operators use the Energy Protocol to dynamically balance energy supply and demand, manage market-based exchanges, and facilitate coordination among local and regional networks.

EnergyNet Operators also maintain key relationships with local grid owners, municipalities, and end-users, ensuring that all parties benefit from a secure, efficient, and highly flexible energy distribution system.

#### 4.12.1 Energy Network Management System: Scalable Operation and Secure Control

As the number of deployed Energy Routers grows, from dozens in a pilot to thousands in a regional network, manual configuration and monitoring quickly becomes infeasible. This is where the EnergyNet Operator steps in, using a purpose-built Energy Network Management System (ENMS) to manage the distributed infrastructure at scale.

The ENMS is the digital control plane for EnergyNet operations. It provides the tools and automation required to provision, monitor, secure, and optimize a dynamic fleet of Energy Routers across neighborhoods, cities, or entire regions.

### 4.12.2 Key Functions of the ENMS

Provisioning & Lifecycle Management

- Remote onboarding of new routers with configuration profiles and Energy Protocol credentials.
- Policy-based role and behavior assignment for each Energy Router depending on its role (e.g., gateway, backbone node, or aggregation point).

Software and Firmware Management

- Secure distribution and orchestration of software updates for EROS and EP-Server.
- Staged rollouts with rollback capability to ensure system stability during upgrades.
- Real-time version tracking and compatibility management across hardware generations.

Advanced Cybersecurity Integration

- Continuous monitoring for anomalies, intrusion attempts, and firmware integrity.
- Role-based access control, encrypted communications, and secure boot validation.
- Integration with national and regional cybersecurity standards and response frameworks.

Predictive Maintenance and Health Monitoring

- Real-time diagnostics on temperature, voltage levels, switching performance, and hardware conditions.
- AI-powered analytics to detect early signs of failure or degradation.
- Predictive scheduling of component replacement or rebalancing of traffic loads.

Coordination and Optimization at Scale

- Real-time visibility into energy flows across the entire mesh of routers.
- Demand forecasting, local load balancing, and market integration.
- Policy enforcement for service level agreements, energy sharing rules, or emergency protocols.

By centralizing control and leveraging automation, the ENMS enables the EnergyNet Operator to offer carrier-grade reliability with the agility of distributed, software-defined infrastructure. ENMS enforces signed software images, mutual-Transport Layer Security (TLS) for EP control traffic, and role-based access (RBAC) with audit trails. Just as Internet Service Providers rely on NMS platforms to manage thousands of routers and switches, the ENMS is the critical software component for safe, scalable, and resilient EnergyNet deployments.

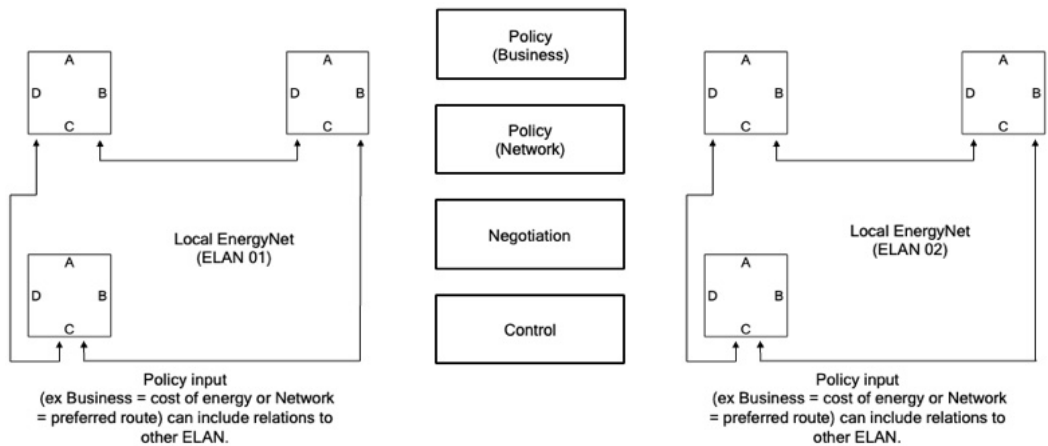

Fig. 8. Illustration of different system levels, priority inside the Energy Router, priority inside the ELAN, and priority between two interconnected ELANs.

### 4.12.3 EnergyNet Operator as a Digital Utility: The BSS/OSS Perspective

To ensure scalable, secure, and efficient deployment and operation of decentralized energy networks, the EnergyNet Operator (ENO) adopts a structure similar to that of modern telecom or cloud service providers. Using the Enhanced Telecom Operations Map (eTOM) model from TM Forum as a reference, the ENO's functions span both Business Support Systems (BSS) and Operational Support Systems (OSS).

This model ensures that the operator can manage customer relationships, maintain network performance, and handle resource orchestration in a modular and standards-based way.

This also enables market competition and interoperability: multiple ENOs could operate over the same physical infrastructure, just like multiple ISPs or Mobile Virtual Network Operators (MVNOs) share telecom networks, encouraging innovation, customer choice, and regulatory oversight.

## 4.13 Neutral Marketplaces Operated by the EnergyNet Operator

In addition to managing energy routing and operational services, the EnergyNet Operator (ENO) plays a key role as a neutral market facilitator, enabling dynamic and rule-based energy transactions across all layers of the EnergyNet architecture.

### 4.13.1 Local Marketplaces (ELAN Level)

Within an Energy LAN (ELAN), such as a neighborhood, residential block, or industrial cluster, the ENO can operate a closed-loop marketplace where:

- Producers (e.g., rooftop solar, battery owners) offer energy capacity or flexibility.
- Consumers (e.g., homes, appliances, heat pumps, EVs) express real-time demand profiles or preferences.
- Routing priorities are negotiated dynamically via the Energy Protocol.

- Free energy sharing (peer-to-peer or community-based peering) can be supported to enable energy solidarity or nonprofit pooling models.

This local energy market functions with millisecond precision and high autonomy, even during upstream grid disconnection, ensuring resilience and resource optimization. Local peering (free sharing) can coexist with priced exchanges; EP messages capture policy/priority so that resilience goals are preserved even during market-based transactions.

### 4.13.2 Regional & Cross-ELAN Trading (EWAN Level)

Across an Energy WAN (EWAN), spanning towns, campuses, or districts, the ENO can:

- Aggregate surplus energy or flexibility from multiple ELANs.
- Operate routing nodes that match supply and demand in real time.
- Allow inter-ELAN energy trading, where energy can be monetized or bartered between cooperating microgrids.
- Set dynamic pricing signals or constraints to reflect network health, urgency, or policy objectives (e.g., carbon intensity).

### 4.13.3 Interconnection to National and Third-Party Markets

Through standard APIs and secure interconnection layers, the ENO can integrate with:

- National flexibility markets, such as those operated by transmission system operators (TSOs).
- Wholesale and balancing markets for ancillary services or frequency control.
- Third-party energy service providers, energy communities, and retail suppliers offering innovative tariffs or grid services.

This ensures full vertical integration of the EnergyNet, from individual homes and buildings all the way up to national and transnational market structures, while preserving local autonomy, microgrid to microgrid interoperability, and traditional grid compatibility.

## 4.14 Energy Resources Integration

EnergyNet leverages diverse local energy resources, seamlessly integrating them to create a resilient and responsive energy network. This integration optimizes the use of renewable energy generation, storage, and flexible load management at the local and regional levels. EnergyNet Operators can facilitate and recommend, though it is always the owners of the energy resources that have final say on management decisions such as price and priority.

### 4.14.1 Local Energy Generation

EnergyNet efficiently integrates local renewable generation, primarily from solar photovoltaic (PV) systems and local wind installations. These distributed resources provide clean renewable energy, significantly reducing dependency on centralized generation and enabling greater local autonomy.

### 4.14.2 Local Energy Storage

A critical component of EnergyNet is energy storage, which balances supply and demand. Storage solutions include:

- Battery Storage Systems (stationary batteries, residential or commercial scale).
- Electric Vehicle (EV) batteries function as mobile storage assets (Vehicle-to-Grid).
- Alternative storage methods, including hydrogen-based storage systems.

### 4.14.3 Local EV Charging / Mobile Storage

Electric vehicle chargers in EnergyNet can do more than simply charge vehicles. Through smart bidirectional chargers, EVs become flexible balancing resources. They can store excess renewable energy and feed it back into local microgrids (Vehicle-to-Grid) when required, stabilizing and optimizing local energy flows.

## 4.15 Physical Infrastructure: "Freedom Cables"

EnergyNet can operate on existing local grid infrastructure, if the grid owner so decides. However, in areas where grid owners have not yet adopted EnergyNet, new infrastructure known as "Freedom Cables" can be deployed. These dedicated parallel cables ensure robust and flexible energy distribution, independently managed from legacy grid constraints. Inside the European Union through the EU's Energy Communities policy, traditional grid owners will no longer be able to legally block construction of new parallel power cable infrastructure [15]. This will make it much easier to bring new innovative energy solutions to the marketplace. It also can accelerate the adaptation of new architectures such as EnergyNet by traditional grid operators.

### 4.15.1 Deployment Conditions and Benefits

"Freedom Cables" can become necessary when traditional grid owners do not promptly adopt the open EnergyNet standard. Deploying these parallel cables enables communities and municipalities to rapidly achieve:

- Energy autonomy and enhanced local resilience.
- Accelerated integration of renewable energy generation and storage solutions.
- Ability to dynamically route energy, bypassing legacy grid bottlenecks.

# 5. Real-World Examples: Key Organizations and Projects

Across the European Union, cities such as Lund and Örebro in Sweden have begun deploying new DC microgrids, creating local demonstrations of EnergyNet's full potential. These pilots validate both technical feasibility and economic benefits, proving that EnergyNet can be scaled quickly, cost-effectively, and reliably.

## 5.1 Viable Cities: Leading Swedish Cities Towards Climate Neutrality

Viable Cities, a strategic innovation program funded by Vinnova, the Swedish Energy Agency, and Formas, is pivotal in driving the climate-neutral transition across Sweden's cities. Uniting 48 municipalities, that together contain more than half the Swedish population, Viable Cities is spearheading local actions aimed at reaching climate neutrality by 2030 [16].

### 5.1.1 From Local System Demonstrators to Global Impact

Viable Cities emphasizes scalable and replicable solutions that can transcend Sweden's borders. By developing and refining effective energy-sharing practices, smart infrastructure, and sustainable mobility systems within Swedish municipalities, Viable Cities creates practical blueprints that other cities can adopt. This approach mirrors Sweden's pioneering telecom deregulation of the 1980s and 1990s, when strategic regulatory shifts and innovation policies demonstrated globally replicable models for market-driven infrastructure investment.

### 5.1.2   Regulatory and Market Synergies

Just as Swedish telecom deregulation sparked competitive innovation and infrastructure transformation, Viable Cities leverages policy frameworks like the EU's energy community directives. Through active engagement in regulatory reform, Viable Cities ensures that cities can effectively implement innovative climate solutions such as EnergyNet-enabled communities and decentralized renewable energy systems.

### 5.1.3   Setting Global Standards

Sweden's early leadership in telecom deregulation positioned it as a global telecommunications innovator. Similarly, Viable Cities positions Sweden as a leading nation in climate-neutral urban transitions. By scaling successful models globally, Viable Cities helps transform climate neutrality from local ambition to global de facto standards, demonstrating that effective regulation, market-driven innovation, and collaborative governance can drive sustainable infrastructure investments worldwide.

## 5.2   Sveriges Allmännytta: Scaling up EnergyNet Across Sweden

Sveriges Allmännytta is the national association representing municipally owned and long-term private housing companies in Sweden, collectively managing over 950,000 apartments across more than 300 housing providers. Together that is more than 20% of all Swedish households. It plays a vital role in advocating for sustainable housing, energy efficiency, and regulatory innovation on behalf of its members.

### 5.2.1   Allmän Energi: From Concept to Collective Action

Under the banner of "Allmän Energi", Sveriges Allmännytta has entered a strategic partnership with ViaEuropa, the world's first EnergyNet Operator, to accelerate the energy transition through municipally owned housing companies [17]. This collaboration aims to streamline the adoption of EnergyNet-based solutions, local energy production, storage, and sharing, across urban housing portfolios.

Housing associations engaged in this initiative include multiple members of Sveriges Allmännytta from places such as Örebro (ÖBO) and Lund (LKF), showcasing the applicability for both existing housing and new development through public-private partnerships implementing citywide deployment plans for each municipality.

## 5.3   Örebro: Blueprint for Energy Communities in Action

The city of Örebro is part of the Viable Cities climate-neutral city alliance, and its municipal housing company, Örebro Bostäder (ÖBO), is a system demonstrator partner for Sveriges Allmännytta's Allmän Energi project, with an initial focus on installing EnergyNet in new developments, through collaboration with multiple partners including private developers and the local grid owner.

### 5.3.1 Tamarinden: A Pioneering Energy Community

Tamarinden, located in Örebro, Sweden, represents an innovative "smart" neighborhood encompassing approximately 800 homes, a nursery, and commercial spaces. Initiated in 2020, Tamarinden stands out by aiming for near-total energy self-sufficiency by integrating solar photovoltaics, geothermal heating, battery storage, and shared electric vehicle infrastructure.

### 5.3.2 Removing Legal Uncertainty for Energy Sharing

Crucially, Tamarinden pioneered the idea of cross-property electricity sharing. Initially restricted by Swedish law, local advocacy and regulatory reform led the Supreme Administrative Court to permit tax-exempt solar energy sharing from 2024 onward. This regulatory breakthrough has positioned Tamarinden as a replicable model for energy communities nationwide.

### 5.3.3 Measurable Impact

Early assessments from Tamarinden highlight significant efficiencies: a 30% reduction in energy consumption and a 50% decrease in peak power demand through predictive and optimized local energy management [18].

## 5.4 Lund: A History of Technology Leadership and Innovation

Lund, a historic university city in southern Sweden, has a long-standing tradition of technology leadership and innovation. Home to Lund University, one of Scandinavia's oldest and most prestigious institutions, the city has been at the forefront of global technological advancements for decades.

The city of Lund is a member of the Viable Cities climate-neutral city alliance. Lund's municipal housing company, Lunds Kommuns Fastighets AB (LKF), is a system demonstrator partner for Sveriges Allmännytta's Allmän Energi project, with an initial focus on installing EnergyNet in existing buildings.

### 5.4.1 GSM and Broadband Leadership

In the 1980s and 1990s, Lund significantly influenced global telecommunications through Ericsson's local research center, and was instrumental in pioneering the GSM (Global System for Mobile Communications) standard. This groundbreaking work laid the foundation for today's global mobile wireless telecommunications connectivity.

In 1998, Lund further cemented its role as a technological innovator when Jonas Birgersson, then at Framfab (based in the IDEON Science Park), developed the world's first commercial broadband service. Officially launched in 1999 as Bredbandsbolaget, this pioneering company quickly became a leading challenger in broadband connectivity, fundamentally changing the landscape of high-speed Internet access and setting the standard for competitive, consumer-driven broadband services worldwide.

### 5.4.2 Innovation Ecosystem: Brunnshög, CoAction Lund, and the birthplace of EnergyNet

Today, Lund continues this tradition of pioneering technological leadership through ambitious initiatives such as the Brunnshög Innovation District, CoAction Lund, and the Lund Green Innovation District.

Brunnshög, located adjacent to world-leading research facilities MAX IV and the European Spallation Source (ESS), is a model of sustainable urban development. The district hosts Sweden's largest low-temperature district heating network, integrating innovative energy solutions, and serving as a real-world testbed for digital energy infrastructure, including EnergyNet-enabled, programmable microgrids, and peer-to-peer energy sharing.

CoAction Lund, a large-scale system demonstrator supported by Vinnova and led by the City of Lund, mobilizes over 25 partners across academia, industry, and government to achieve a climate-neutral city by 2030. CoAction Lund exemplifies how Lund leverages interdisciplinary collaboration, combining research excellence from Lund University, cutting-edge technologies from local companies, and innovative governance, to drive large-scale sustainability transitions. The Lund Green Innovation District is a "Sustainable Innovation Zone" in collaboration with the EnergyNet Task Force, Viable Cities, Global Urban Development (GUD), and other partners [19].

### 5.4.3 Lund University: Engine of Innovation

Lund University, with its strong research in technology, sustainability, and digital infrastructure, has been a catalyst for continuous innovation.

### 5.4.4 Local Leadership

The City of Lund has among its elected officials not only leaders that have the will to continue to build on its history of innovation but also some key assets. With LKF, a for-profit housing and commercial development and management company completely owned by the city government, and the second largest local public utility company in Sweden, Kraftringen AB, that is majority owned by the city government, elected officials have substantial capability to modernize the city's infrastructure and facilitate major new green innovations such as EnergyNet.

### 5.4.5 EnergyNet Technical Proof of Concept

On April 26, 2025, CoAction Lund officially launched the world's first operational EnergyNet system [20] marking a significant global milestone, similar to the first nodes getting connected to the Internet three decades ago. Developed collaboratively by LKF together with the Lunds Kommuns Parkerings AB (LKP) Lund Parking Company and ViaEuropa, this pilot employs an innovative "Freedom Cable" enabling a parallel DC microgrid to provide direct energy sharing between two buildings, each with rooftop solar, located inside both the Brunnshög Innovation District and the larger Lund Green Innovation District.

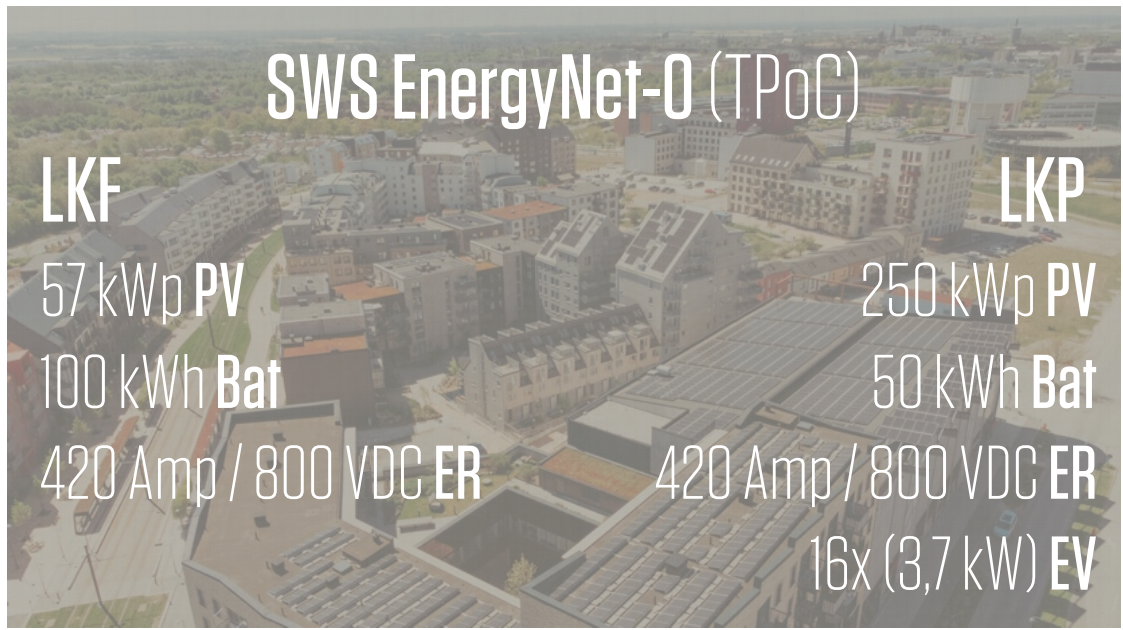

Fig. 9. The official full name of the installation: SWS (Solar-Wind-Storage) EnergyNet System – 0, this illustration shows the capabilities on each side, and the connection.

## 6. Conclusion: From Possibility to Deployment

The traditional grid has brought electrification to billions and powered a century of growth, but it was never designed for decentralization, bidirectionality, or fully mobilizing the revolutionary capabilities of modern electronics and software. What the Internet did for communication, scaling up global access through open standards, hyperscale electronics, and new physical infrastructure, EnergyNet can now do for electricity.

*What's needed is coordination, leadership, and the courage to transform the existing energy distribution system.*

EnergyNet is a path forward that can lead to:

- Low fixed costs for clean local energy.

- Increased local and national energy independence.

- Enhanced resilience with no single point of failure.

- A potential breakthrough in electric vehicle and battery integration.

- An opportunity to free up grid capacity for electrification of data centers and advanced manufacturing facilities.

- Opening up a new era of digitally driven energy innovation.

This paper has outlined the architectural blueprint and operational model needed to move from pilot to scale. Just as countries that led broadband deployment now reap digital dividends, those who move first on EnergyNet can gain competitive energy, economic, and climate advantages that are both durable and transformative.

The grid of the past (and present) is centralized, locked, and fragile. The grid of the future will be distributed, open, and adaptive.

**EnergyNet as the key architecture for the 21st-century grid is an innovative solution that will unlock extraordinary opportunities for the world to "get richer by becoming greener".**

**The future of energy belongs not to those who wait, but to those who act.**

# 8. APPENDIX: What about Smart Grids?

The smart grid era began with high hopes [21]: smart meters, sensors, and upgraded SCADA/ADMS would give operators real-time visibility, finer control, and fewer outages. Those programs delivered real value, better data, faster restoration, and volt/VAR optimization, but they largely left the core architecture unchanged: with radial feeders, one big synchronous machine, and a central control plane polling thousands of endpoints.

As distributed PV, batteries, heat pumps, and EVs surged at the edge, the returns to "more intelligence in the center" began to diminish; interconnection queues, feeder limits, and resilience gaps persisted. David Isenberg and the Net-Head´s telecom lesson applies: the Intelligent Network (smart in the core) could not match the Internet's "stupid network", a simple, abundant core with open Internet and edge intelligence.

Smart grid is the power sector analogue of the "Intelligent Network" in telecommunications: useful, but incremental [22]. Smart grid programs improved data and visibility, but left the radial, synchronous architecture unchanged. EnergyNet changes the architecture through digitally firewalled domains, near-real-time operation with local buffering, and policy-based interconnection.

EnergyNet is the architectural successor: a network of networks with digitally negotiated and galvanically separated borders, software-defined and edge-first control (EROS/EP-Server operated via independent ENMS), near-real-time operation with local buffering ("some power is better than no power"), open protocols such as Energy Protocol (EP) and neutral marketplaces, and port-by-port scaling on demand. **If implemented at scale, EnergyNet has the potential to unlock unlimited green energy abundance.**

| Aspect | "Smart Grid" (incremental) | EnergyNet (architectural) |
|---|---|---|
| Core topology | Same radial, synchronous AC | **Network-of-networks** with **negotiated, galvanic borders** |
| Control | Central SCADA/ADMS/DERMS | **Edge autonomy** (Energy Protocol (EP) + EROS/EP-Server) + fleet ops (independent ENMS) |
| Flow model | Continuous, strict real-time | **Near-real-time** with **local buffering/storage** |
| Interop/markets | Utility platforms, vendor silos | **Open protocol**, neutral **local/regional** marketplaces |
| Scaling | Big, slow reinforcements | **Port-by-port**, **rack-by-rack** modularity, where new capacity is added on demand |

Fig. 10. Smart Grid versus EnergyNet, compared in five key aspects.